\documentclass[a4paper,fleqn]{cas-dc}

\usepackage[numbers]{natbib}

\def\tsc#1{\csdef{#1}{\textsc{\lowercase{#1}}\xspace}}
\tsc{WGM}
\tsc{QE}
\tsc{EP}
\tsc{PMS}
\tsc{BEC}
\tsc{DE}

\begin{document}

\shorttitle{Belle II VTX upgrade}    

\shortauthors{V. Prasad {\it et al.}}  

\title[mode = title]{Upgrade of the Belle II Vertex Detector with Depleted Monolithic
CMOS Active Pixel Sensors}

\author[18]{V.~ Prasad}[ orcid=0000-0001-7395-2318]
\ead{vindy@jlu.edu.cn}
\author[1]{M.~Babeluk}
\author[1]{T.~Bergauer}
\author[1]{M.~Friedl}
\author[1]{C.~Irmler}
\author[1]{B.~Pilsl}
\author[1]{C.~Schwanda}

\affiliation[1]{Marietta Blau Institute for Particle Physics, Austrian Academy of Sciences, Dominikanerbastei 16, 1010 Vienna, Austria}

\author[2]{L.~Gaioni}
\author[2]{V.~Re}
\author[2]{E.~Riceputi}
\author[2]{G.~Traversi}

\affiliation[2]{Department of Engineering and Applied Sciences, University of Bergamo, Viale Marconi 5, I-24044 Dalmine (BG), Italy}

\author[3]{S.~Giroletti}
\author[3]{L.~Ratti}

\affiliation[3]{Department of Electrical, Computer and Biomedical Engineering, University of Pavia, Via Ferrata 5, I-27100 Pavia, Italy}

\author[4,5]{G.~F.~Benfratello}
\author[4,5]{S.~Bettarini}
\author[5]{F.~Bosi}
\author[4,5]{G.~Casarosa}
\author[5]{L.~Corona}
\author[4,5]{F.~Forti}
\author[5]{A.~Gabrielli}
\author[5]{M.~Massa}
\author[4,5]{L.~Massaccesi}
\author[5]{M.~Minuti}
\author[5]{A.~Moggi}
\author[4,5]{S.~Mondal}
\author[4,5]{G.~Rizzo}
\author[5]{M.~Rovini}
\author[5]{A.~Taffara}

\affiliation[4]{Dipartimento di Fisica “E. Fermi”, Universit\`a di Pisa, L.go B. Pontecorvo 3, I-56127 Pisa, Italy}
\affiliation[5]{INFN Sezione di Pisa, L.go B. Pontecorvo 3, I-56127 Pisa, Italy}

\author[6]{M.~Barbero}
\author[6]{P.~Barrillon}
\author[6]{R.~Boudagga}
\author[6]{P.~Breugnon}
\author[6]{D.~Fougeron}
\author[6]{P.~Pangaud}
\author[6]{J.~Serrano}
\author[6]{V.~Vobbilisetti}
\author[6]{D.~Xu}

\affiliation[6]{Aix Marseille Univ, CNRS IN2P3, CPPM, Marseille,  France}

\author[7]{D.~Auguste}
\author[7]{J.~Bonis}
\author[7]{Y.~Peinaud}
\author[7]{M.~Winter}

\affiliation[7]{Laboratoire de Physique des 2 infinis Ir\`ene Joliot-Curie – IJCLab, Université Paris-Saclay, CNRS IN2P3, IJCLab, 91405 Orsay, France}

\author[8]{J.~Baudot}
\author[8]{G.~Bertolone}
\author[8]{A.~Dorokhov}
\author[8]{G.~Dujany}
\author[8]{L.~Federici}
\author[8]{C.~Finck}
\author[8]{A.~Himmi}
\author[8]{C.~Hu-Guo}
\author[8]{A.~Kumar}
\author[8]{M.~Maushart}
\author[8]{F.~Morel}
\author[8]{H.~Pham}
\author[8]{I.~Ripp-Baudot}
\author[8]{R.~Sefri}
\author[8]{P.~Stavroulakis}
\author[8]{I.~Valin}

\affiliation[8]{Universit\'e de Strasbourg, CNRS, IPHC UMR 7178, F-67000 Strasbourg, France}

\author[9]{F.~Bernlochner}
\author[9]{C.~Bespin}
\author[9]{J.~Dingfelder}
\author[9]{T.~Kishishita}
\author[9]{H.~Krüger}
\author[9]{L.~Schall}
\author[9]{M.~Vogt}

\affiliation[9]{Physikalisches Institut, Rheinische Friedrich-Wilhelms-Universität Universität Bonn, Nussallee 12, 53115 Bonn, Germany}

\author[10]{M.~Karagounis}

\affiliation[10]{University of Applied Sciences and Arts Dortmund, Sonnenstraße 96-100, 44139 Dortmund, Germany}

\author[11]{Y.~Buch}
\author[11]{A.~Frey}
\author[11]{B.~Schwenker}
\author[11]{M.~Schwickardi}

\affiliation[11]{II. Physikalisches Institut, Georg-August-Universität Göttingen, Friedrich-Hund-Platz 1, 37077 Göttingen, Germany}

\author[12,13]{K.~Hara}
\author[12,13]{D.~Jeans}
\author[12,13]{K.~R.~Nakamura}
\author[12,13]{Y.~Okazaki}

\affiliation[12]{High Energy Accelerator Research Organization (KEK), Tsukuba 305-0801, Japan}
\affiliation[13]{The Graduate University for Advanced Studies (SOKENDAI), Hayama 240-0193, Japan}

\author[14]{T.~Higuchi}

\affiliation[14]{Kavli Institute for the Physics and Mathematics of the Universe (WPI), University of Tokyo, Kashiwa-no-ha 5-1-5, Kashiwa 277-8583, Japan}

\author[15]{Y.~Onuki}
\author[15]{S.~Wang}

\affiliation[15]{Department of Physics, University of Tokyo, Hongo 7-3-1, Tokyo 113-0033, Japan}

\author[16]{C.~Lacasta}
\author[16]{C.~Marinas}
\author[16]{J.~Mazorra de Cos}
\author[16]{L.~Molina-Bueno}

\affiliation[16]{Instituto de Fisica Corpuscular (IFIC), CSIC-UV, Catedratico Jose Beltran, 2. E-46980 Paterna, Spain}

\author[17]{A.~Bevan}
\author[17]{M.~Bona}
\author[17]{D.~Howgill}

\affiliation[17]{School of Physical and Chemical Sciences, Department of Physics and Astronomy, Queen Mary University of London, 327 Mile End Road, London, E1 4NS, United Kingdom}

\author[18]{W.~ Song}
\author[18]{J.~Gong}
\author[18]{X.~Gao}

\affiliation[18]{College of Physics, Jilin University
, 2699 Qianjin Street, Changchun, Jilin, China}

\author[19]{A.~Fernandez~Prieto}
\author[19]{A.~Gallas~Torreira}

\affiliation[19]{Universidade de Santiago de Compostela, 2010 Instituto Galego de Física de Altas Enerxías (IGFAE), Colexio de San Xerome, PZ Obradoiro,  S N. E-15782 Santiago de Compostela, Spain}




\begin{abstract}
The Belle II experiment, operating at the asymmetric SuperKEKB $e^+e^-$ collider, is preparing an upgrade of its vertex detector to cope with an increased luminosity of $6 \times 10^{35}$ cm$^{-2}$s$^{-1}$. The upgraded vertex detector (VTX) will consist of five or six layers of depleted monolithic active pixel sensors (DMAPS), with a total material budget of approximately $3\%$  $X/X_0$.  The OBELIX chip, derived from the TJ-Monopix2 sensor and fabricated using Tower-Jazz  Semiconductor 180 nm CMOS technology, is being developed for this upgrade. It features a 33 $\mu$m pixel pitch with a time-stamping binning of $50-100$ ns, along with a dedicated digital periphery compatible with the Belle II trigger system, supporting rates up to 30 kHz. The sensor is designed to operate under the high background conditions expected at the target luminosity, with radiation tolerance up to $5 \times 10^{14}$ $n_{eq}$/cm$^2$ and 100 Mrad, while targeting a power density of about 200 mW/cm$^2$. This corresponds to hit rates up to 120 MHz/cm$^2$. Beam test and irradiation studies of TJ-Monopix2 demonstrate that the operating sensor temperature should stay below   $40^\circ$C after irradiation up to $5 \times 10^{14}$ $n_{eq}$/cm$^2$. This report reviews the proposed VTX concept, sensor performance, and ongoing R$\&$D activities.

\end{abstract}


\maketitle

\section{Introduction}
Belle II~\cite{Belle2} is a second-generation B-factory experiment operating at the asymmetric SuperKEKB $e^+e^-$ collider in Tsukuba, Japan~\cite{superKEKB}. Its predecessors, the BaBar and Belle experiments, established the Kobayashi–Maskawa mechanism of $CP$ violation, which led to the Nobel Prize in Physics in 2008~\cite{BFactory}. The experiment provides asymmetric collisions of 7 GeV electrons and 4 GeV positrons, tuned to the $\Upsilon(4S)$ resonance. Belle II aims to accumulate an integrated luminosity approximately 50 times larger than that of its predecessor, using high beam currents and the nano-beam scheme with a cost of challenging background conditions~\cite{Belle2upgrade}. To handle the resulting increase in event rates and large backgrounds, most of its subdetector components are either redesigned or completely replaced.

To achieve precise track reconstruction in the vicinity of the interaction point, the current vertex detector (VXD) is composed of six layers of silicon-based sensors. The VXD design consists of two layers of DEPFET-based pixel detectors (PXD)~\cite{pxd} and four layers of silicon strip detectors (SVD)~\cite{svd}.  The PXD layers are located closest to the interaction point and feature a pixel pitch of $50\text{--}70~\mu\mathrm{m}$ with an integration time of $20~\mu$s. The SVD is composed of double-sided silicon strip detectors, with strip lengths of up to $6~\mathrm{cm}$ and a time resolution of $20~\mathrm{ns}$.  For particles at normal incidence, the material budget per layer varies from $0.25\%$ to $0.75\%$ of radiation length, leading to an overall material budget of approximately $3.5\%~X/X_0$. The VXD offers spatial resolution between 10 and 25 $\mu\mathrm{m}$, with coverage over a polar angle range of $17\text{--}150$ degrees, spanning radii from 14 mm to 140 mm.

Belle II began physics data taking in 2019 and has accumulated approximately 400~fb$^{-1}$ of Run~1 data, leading to more than 90 peer-reviewed publications. During the long shutdown 1 (LS1), several upgrades were implemented, including the installation of new collimators, the second‑layer PXD, and improvements to the interaction region (IR). Data taking for Run~2 has been ongoing since 2024. The peak luminosity has already reached $5.2 \times 10^{34}$ cm$^{-2}$s$^{-1}$, and further aim is to achieve the design target luminosity of $6.0 \times 10^{35}$ cm$^{-2}$s$^{-1}$~\cite{Belle2upgrade}. However, within the constraints of the current accelerator configuration, SuperKEKB is expected to achieve only about $2.0 \times 10^{35}$~cm$^{-2}$s$^{-1}$~\cite{Belle2upgrade}.

Achieving the target peak luminosity will require significant upgrades to the accelerator complex, potentially including a redesign of the IR, which could impact the VXD envelope. The current VXD provides excellent tracking capabilities but has limitations in handling the very high-background rates expected from extrapolated beam backgrounds, which could reach approximately $2\%$ occupancy in the innermost PXD layer (PXD layer~1) and $8\%$  occupancy in SVD layer~3, potentially degrading the tracking performance and the overall robustness of the detector~\cite{Belle2upgrade}. Under these conditions, a new vertex detector with higher spatial and time granularity is desirable to preserve tracking and vertexing capabilities. An LS2 is planned around 2032 for a major upgrade and redesign of the IR region and   superconducting final focus. It also provides an ideal opportunity for the vertex detector upgrade. Due to uncertainties in background predictions and the accelerator redesign, a new robust vertex detector (VTX) is proposed, as described below and in ref.~\cite{Belle2upgrade}.

\section{Upgrade requirements and VTX proposal}
The Conceptual Design Report~\cite{Belle2upgrade} investigates a set of key performance requirements to address the physics goals of Belle II at the target luminosity. These requirements include:

\begin{itemize}
\item enhanced spatial resolution and granularity to cope with severe beam-induced backgrounds,
\item pixel pitches in the range of $30-40~\mu$m,
\item spatial resolution better than 15~$\mu \mathrm{m}$,
\item a total material budget of approximately $3\%$ of radiation length,
\item hit-rate capability up to $120~\mathrm{MHz/cm^2}$,
\item radiation tolerance to a Total Ionizing Dose (TID) up to $100~\mathrm{Mrad}$ and $5\times10^{14}~n_{\mathrm{eq}}/\mathrm{cm}^2$ Non-Ionizing Energy Loss (NIEL) for the innermost layer,
\item power dissipation of about $200~\mathrm{mW/cm^2}$ to limit the cooling material.
\end{itemize}

\noindent The technology chosen to meet these requirements is based on Depleted Monolithic Active Pixel Sensors (DMAPS), originally developed for the outer layers of the ATLAS ITk detector~\cite{Atlas}, and in particular the TJ‑Monopix2 sensor~\cite{TJMonopix}, produced using the TowerJazz 180 nm CMOS process.
 The proposed upgrade will replace the existing PXD and SVD with a fully pixelated VTX detector~\cite{Belle2upgrade}.

\section{VTX baseline}
\begin{figure}
	\centering
	\includegraphics[width=1.0\columnwidth]{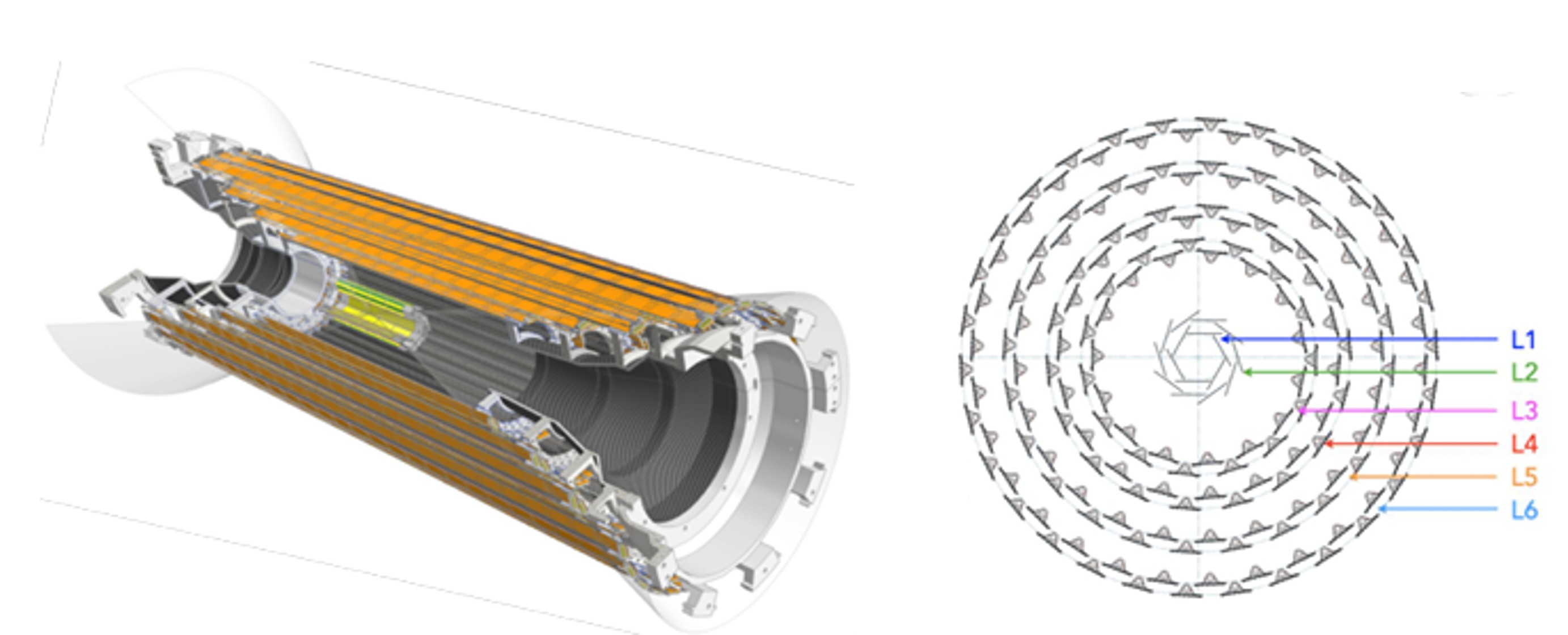}
	\caption{Schematic view of the VTX baseline layout with six detection layers. }
	\label{FIG:1}
\end{figure}

 The baseline design of the new VTX features five or six straight detection layers. All layers are equipped with a common DMAPS monolithic pixel sensor, the Optimized Belle II Monolithic Pixel Sensor (OBELIX) chip~\cite{Obelix}, derived from the TJ-Monopix2 design. This configuration prioritizes improved timing and spatial granularity while maintaining a low material budget compared to the current VXD. Figure~\ref{FIG:1} shows a schematic view of the baseline VTX design with six detection layers. The integration of DMAPS technology results in significant improvements in tracking and impact parameter resolution, as demonstrated by dedicated Monte Carlo simulations described in Ref.~\cite{MC-simulation}. The VTX layers below 30 mm are grouped into the so-called inner VTX (iVTX), while the layers at outer radii are referred to as the outer VTX (oVTX).

\subsection{Inner VTX layers (iVTX)}
\begin{figure}
	\centering
	\includegraphics[width=1.0\columnwidth]{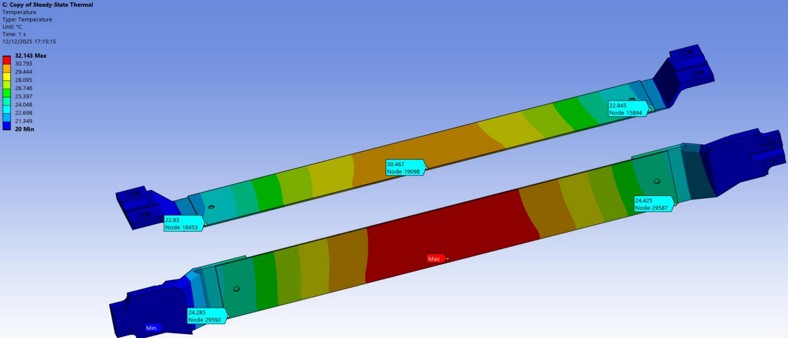}
	\caption{Simulation of the temperature gradient reached along the structure of the iVTX ladder. }
	\label{FIG:2}
\end{figure}

The inner vertex detector (iVTX) comprises the two innermost layers, located at radii of $14~\mathrm{mm}$ and $22~\mathrm{mm}$, and employs a silicon-based ladder concept, with each ladder having an active length of $120~\mathrm{mm}$. Each ladder consists of  silicon pieces comprising four adjacent OBELIX chips~\cite{Obelix} from the same wafer, interconnected via a post-processed redistribution layer (RDL) and potentially thinned to approximately 50 $\mu$m. The target material budget is approximately $0.3\%~X/X_0$ per layer. From the OBELIX design advances, the power dissipation in the iVTX is expected to reach up to $200~\mathrm{mW/cm^2}$. Several cooling options are being investigated to manage the high power density and the degradation of chip performance with increasing temperature after irradiation.  
One such option under investigation is a passive conduction-cooling concept based on a Parylene-coated thermal pyrolytic graphite (TPG) support~\cite{TPG} and dry nitrogen for humidity control. The design goal is to maintain the sensor operating temperature below $40^\circ$C. Fig.~\ref{FIG:2} shows a simulation of the temperature gradient reached along the structure of the iVTX ladder.

Studies of TJ-Monopix2, as described in Section~\ref{TJmonopix}, play an important role in defining the operational temperature limits of the OBELIX sensor. In particular, measurements of irradiated TJ-Monopix2 chips at elevated temperatures provide essential input for evaluating performance degradation and determining safe operating conditions. These results provide important input for the thermal design and cooling strategy of the iVTX.

\subsection{Outer Vertex Detector}
\begin{figure}
	\centering
	\includegraphics[width=1.0\columnwidth]{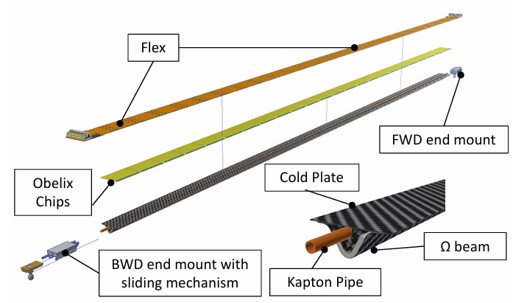}
	\caption{The schematic view of oVTX. }
	\label{FIG:3}
\end{figure}

The outer vertex detector (oVTX) consists of three or four additional layers extending to radii up to $140~\mathrm{mm}$. The ladder concept is inspired by the ALICE ITS~\cite{Alice} design and employs an omega-shaped carbon-fiber support structure with Airex foam core,  a cold plate and an embedded water-cooling Kapton pipe.  Flex circuits connect up to 12 OBELIX~\cite{Obelix} chips to a module, which is supplied and read out from the end of the ladder.  The material budget is approximately $0.6\%~X/X_0$ per layer. A $70~\mathrm{cm}$ long prototype ladder has been mechanically and thermally characterized, demonstrating performance well within specifications~\cite{oVTXprot}. The schematic view of the oVTX is shown in Fig.~\ref{FIG:3}.

\section{TJ-Monopix2 as a Prototype Sensor}
\label{TJmonopix}
\begin{figure}
	\centering
	\includegraphics[width=0.6\columnwidth]{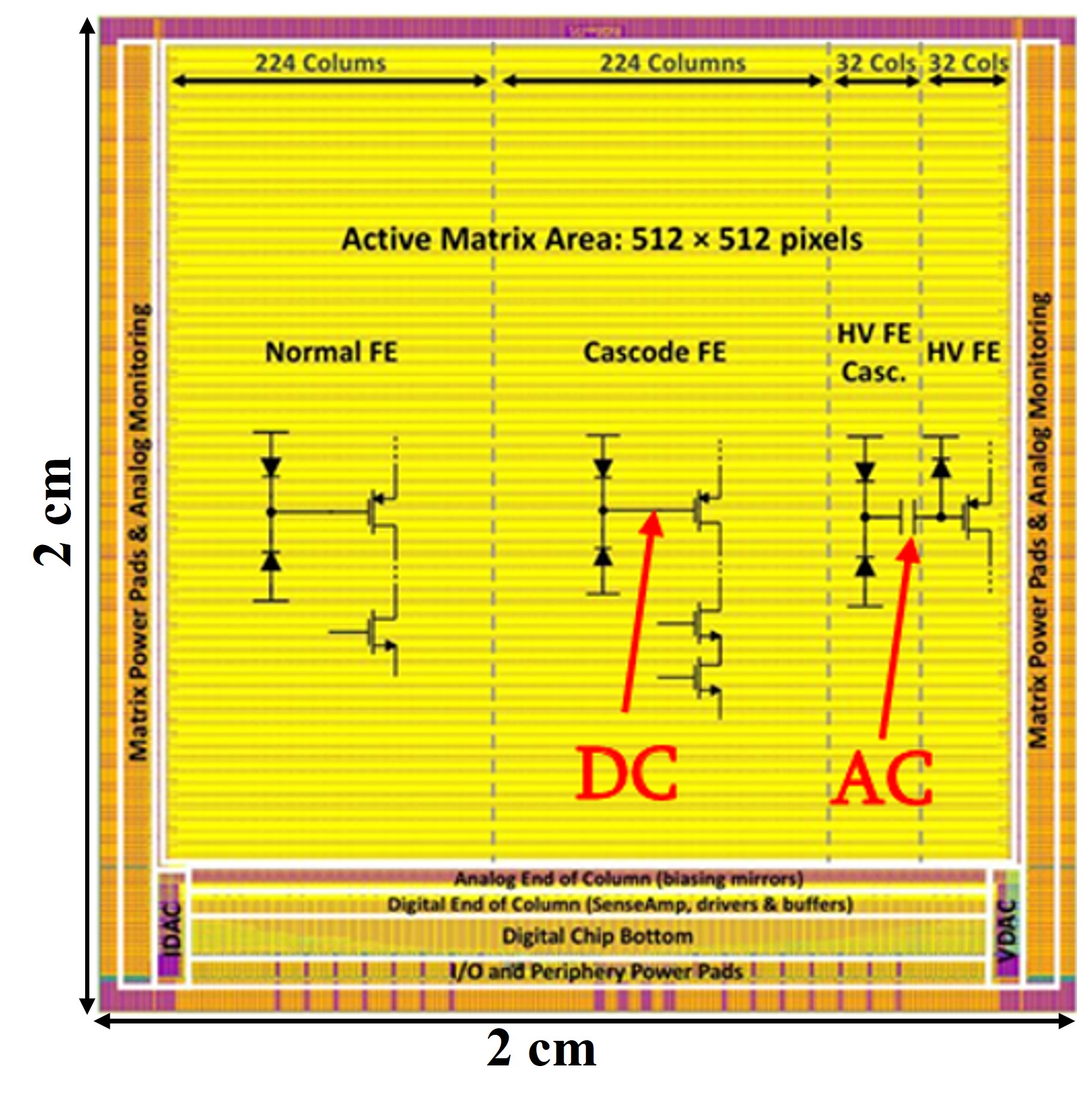}
	\caption{The layout of the $2\times 2$ TJ-Monopix2 CMOS pixel sensor. }
	\label{FIG:4}
\end{figure}

The OBELIX chip for the Belle II VTX upgrade is based on the TJ-Monopix2 CMOS sensor~\cite{TJMonopix}, which is built using Tower-Jazz 180 nm technology with a modified process to improve radiation hardness~\cite{radhard}. The TJ-Monopix2~\cite{TJMonopix} features a square pixel pitch of $33.04\ \mu\rm{m}$, arranged in a $512 \times 512$ matrix covering an active area of approximately $2\times2\ \mathrm{cm}^2$, as shown in Fig.~\ref{FIG:4}. It provides 25 ns hit timestamping and delivers analog Time-over-Threshold (ToT) information with 7-bit precision.  The chip incorporates four front-end (FE) variants, including both DC- and AC-coupled designs. Two of these variants employ a cascode transistor in the voltage amplifier stage to enhance performance. The readout is based on a column-drain architecture, enabling the device to sustain hit rates above 100 MHz/cm$^2$.

TJ-Monopix operates in a triggerless readout mode and lacks peripheral memory and logic for handling trigger signals. The readout provides 7‑bit time‑over‑threshold information and uses the BDAQ53 data acquisition system based on the RD53 architecture~\cite{RD54}.
In contrast, OBELIX integrates a dedicated digital periphery designed to meet the requirements of Belle II, enabling triggered operation, latency buffering, and additional functionalities such as track-trigger information and high-resolution time reconstruction~\cite{Obelix}.

Radiation tolerance of TJ-Monopix2 has previously been demonstrated at low operating temperatures~\cite{TJ-Monopix-temp}. Current tests aim to demonstrate radiation tolerance at room temperature, which is essential for the Belle~II vertex detector upgrade.

The test-beam measurement is based on a $4.2~\mathrm{GeV}$ electron beam provided by the DESY~II ring~\cite{testbeam}. The telescope consists of six layers of MIMOSA26 (three upstream and three downstream) and a TelePix sensor for precise timing measurements. A dedicated cooling setup employing a copper plate in thermal contact with a Peltier element has been used to control the sensor temperature. The operating temperature, measured by an on-chip NTC sensor ($T_{\mathrm{NTC}}$), is varied between $10^\circ$C and $50^\circ$C. The measured $T_{\mathrm{NTC}}$ is approximately $7^\circ$C lower than the actual sensor temperature ($T_{\mathrm{sensor}}$), due to an opening in the copper plate at the chip position.

\begin{figure}
	\centering
	\includegraphics[width=1.02\columnwidth]{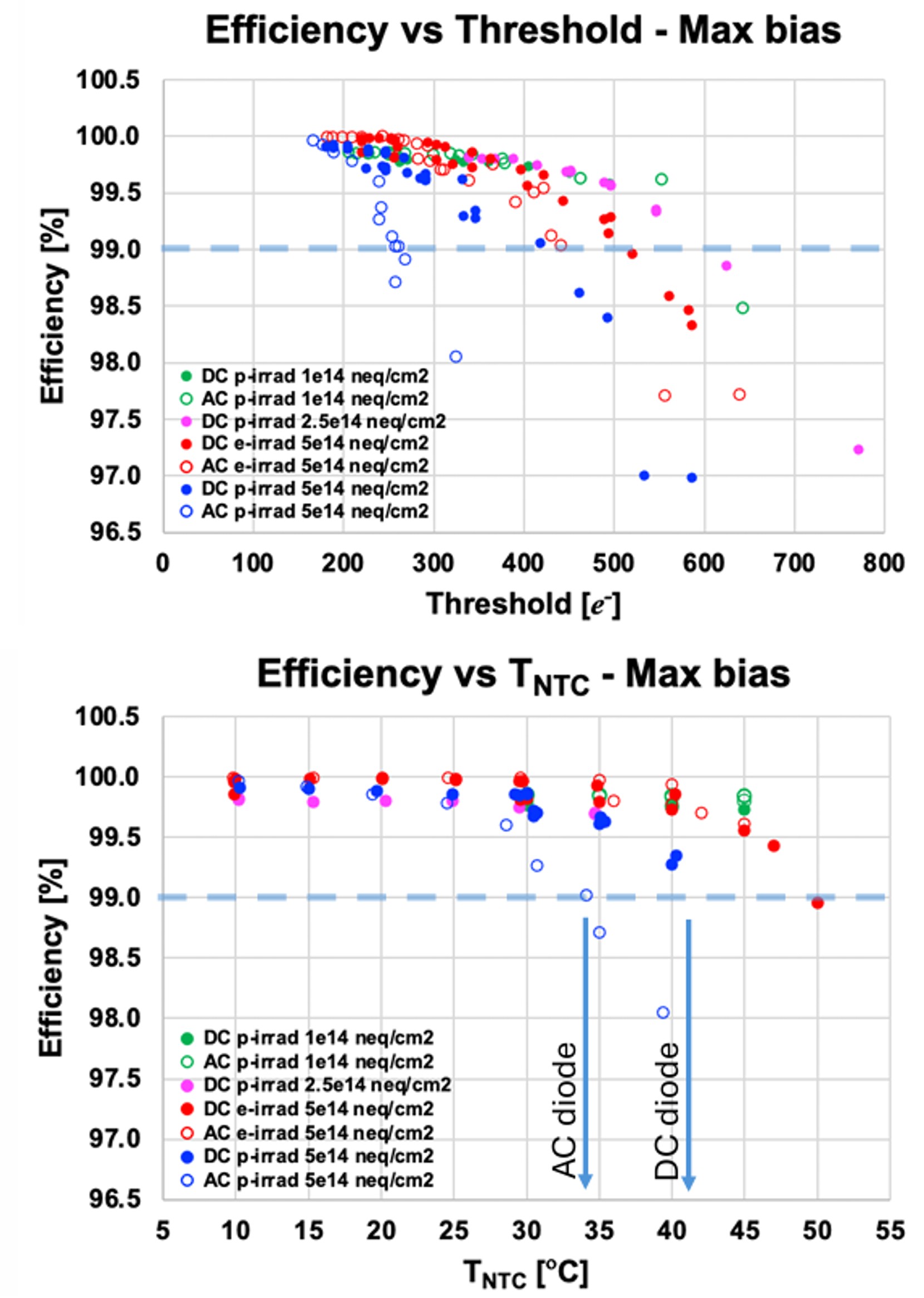}
	\caption{(Top) Efficiency versus $e^-$ threshold and (bottom) efficiency versus $T_{\mathrm{NTC}}$ for different radiation sensors and both DC-coupled (filled dots) and AC-coupled (empty dots) frontends. }
	\label{FIG:5}
\end{figure}

Figure~\ref{FIG:5} shows the hit detection efficiency of TJ-Monopix2 measured at maximum bias voltage after irradiation, as a function of the discriminator threshold (top) and the sensor temperature (bottom). Measurements are shown for both DC- and AC-coupled pixel variants irradiated with either 24 MeV protons or 90 MeV electrons, reaching NIEL fluences of up to  $5\times10^{14}~n_{\mathrm{eq}}/\mathrm{cm}^2$. The efficiency as a function of threshold exhibits the expected degradation with increasing irradiation fluence.

For DC-coupled pixels, efficiencies above $99\%$ are maintained up to thresholds of approximately $400~\mathrm{e^-}$, whereas AC-coupled pixels show a steeper efficiency loss, dropping below $99\%$ at thresholds around $250~\mathrm{e^-}$. Overall, the DC-coupled design demonstrates superior robustness against radiation-induced signal degradation. Although both coupling schemes achieve efficiencies close to $99\%$ at low temperatures, the AC-coupled pixels exhibit a more pronounced efficiency loss at elevated temperatures. In contrast, DC-coupled pixels maintain stable performance up to temperatures of approximately $40^\circ$C. These results indicate that, after irradiation to $5\times10^{14}~n_{\mathrm{eq}}/\mathrm{cm}^2$, the operating temperature should be kept below $40^\circ$C to ensure an efficiency above $99\%$, supporting the DC-coupled pixel architecture as the baseline choice for the Belle~II vertex detector upgrade.

\section{OBELIX chip}
\begin{figure}
	\centering
	\includegraphics[width=0.9\columnwidth]{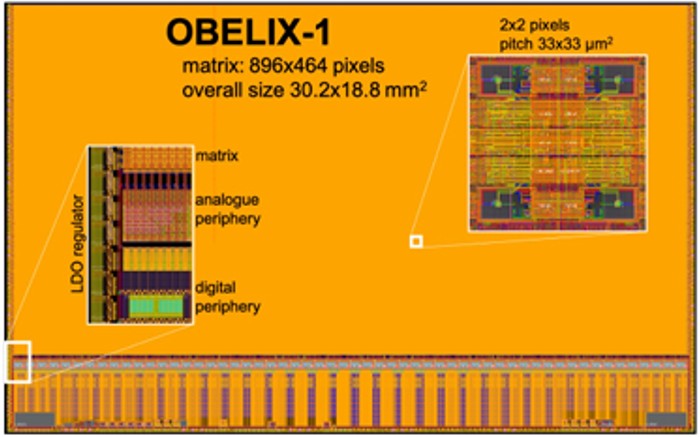}
	\caption{Floorplan illustration of the OBELIX chip. }
	\label{FIG:6}
\end{figure}

OBELIX is an acronym of Optimized Belle II Monolithic Pixel sesor proposed as the main option to equip all VTX layers~\cite{Obelix}. The chip size is $30 \times 19$~$\mathrm{mm^2}$, with an expanded matrix of 464 rows and 896 columns. It  provides hit time-stamping with $50-100$ ns resolution, a 7-bit  ToT readout, and 3-bit local threshold adjustment to mitigate dispersion across the matrix. The floor plan and dimensions of the chip are shown in Fig.~\ref{FIG:6}.
 To simplify the powering scheme over the ladders, on-chip low-dropout regulators (LDOs) are implemented to provide local voltage regulation.  A custom digital periphery has been introduced to meet the specific requirements of the Belle II environment, including a triggered readout scheme managed by the Trigger Readout Unit (TRU), supporting average hit rates up to $120~\mathrm{MHz/cm^2}$, peak tolerance reaching up to $600~\mathrm{MHz/cm^2}$, and trigger rates of $30~\mathrm{kHz}$ with latencies up to $10~\mu$s.

Additional functionality is provided by two dedicated modules in the digital periphery. The Track Trigger Transmission (TTT) module delivers low-latency hit data to the trigger system, enabling fast track-finding in FPGA-based processors with a latency of approximately 200 ns, primarily using hits from the outer VTX layers due to occupancy constraints in the inner layers. The Periphery Time-to-Digital (PTD) module enhances timing precision in low-rate regions by exploiting the column-wise HitOr signal sampled at a higher clock frequency, achieving a timing resolution of 5 ns after corrections. This approach is effective in outer layers where hit rates are below $10~\mathrm{MHz/cm^2}$. The OBELIX sensor submission is taking place in April 2026, and the sensors are expected to be returned in September~2026.

\section{Inner Timing Tracker Concept}

The proposed upgrade of the Belle~II will also include a replacement of the central drift chamber (CDC), which will exclude the inner layers. This introduces a gap  between the outermost VTX layers and the innermost CDC superlayer, leaving tracks traversing this transition region only partially unmeasured. This gap in track coverage will primarily affect low-momentum particles,  for which multiple scattering and limited lever arm can degrade tracking performance, including track finding efficiency and momentum resolution. It will also have a significant impact  on the reconstruction of long-lived particles.  

To mitigate this issue, the introduction of an Inner Timing Tracker (ITT) between the VTX and the CDC is under consideration. The primary physics goals of the ITT are to improve tracking efficiency and robustness in the VTX--CDC transition region and to provide additional particle identification capability, especially for pion--muon separation at low momentum.

\begin{figure}
	\centering
	\includegraphics[width=1.0\columnwidth]{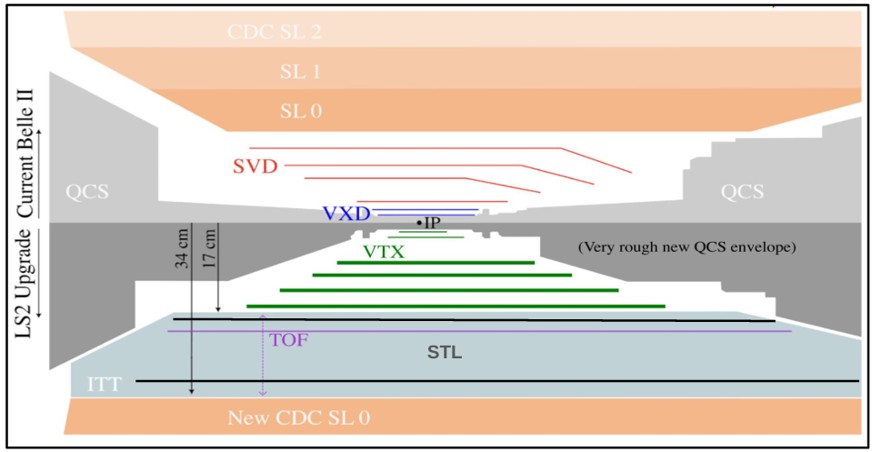}
	\caption{ Conceptual layout of the ITT and its integration with the Belle~II detector. }
	\label{FIG:7}
\end{figure}

Several technology options are currently being investigated. One approach is a Fast Timing Layer (FTL) consisting of a single layer of AC-LGAD sensors to provide precise timing information for improved background rejection and particle identification. Another option is the addition of a Silicon Transition Layer (STL) consisting of two double layers of CMOS strip sensors, providing high-precision spatial measurements to ensure tracking continuity between the VTX and the CDC. The conceptual layout of the ITT and its integration within the Belle~II detector are shown in Fig.~\ref{FIG:7}. Further studies are ongoing to optimize the detector technology and geometry.

\section*{Summary}
The Belle II VTX upgrade is motivated by high background occupancy, radiation damage, and the need for safe operation to achieve target luminosity. A DMAPS-based fully pixelated VTX consists of iVTX and oVTX, satisfying the required performance. The iVTX employs a self-supporting silicon ladder with RDL routing and a high thermal conductivity sheet for cooling, while the oVTX is based on an omega-shaped carbon-fiber structure with Airex foam as core material. Results from TJ-Monopix2~\cite{TJMonopix} strongly support the OBELIX sensor design. OBELIX integrates multiple on-chip functionalities, including fast trigger logic, precise timing, and high-speed data transmission. The R$\&$D of an ITT between VTX and CDC is also under consideration to improve the tracking efficiency of low-momentum tracks and achieve better pion–muon separation.

\section{Acknowledgements}
This work has received the support from  the European Union’s Horizon 2020 Research and Innovation programme under Grant Agreements no 101004761 (AIDAinnova),  Multilateral Scientific and Technological Cooperation in the Danube Region (MULT 03/23),
Horizon 2020 ERC-Consolidator Grant No. 819127,
TY-FJPPN (Toshiko Yuasa-France Japan Particle Physics Network), the MCIU with funding from the European Union NextGenerationEU (PRTR-C17.I01), Seed funding of Jilin University, and Generalitat Valenciana (GVANEXT),  Project ASFAE/2022/016. The measurements leading to these results have been performed at the Test Beam Facility at DESY Hamburg (Germany), a member of the Helmholtz Association (HGF), and received support from the CYRC\'e member of the France Life Imaging network (ANR-11-INBS-0006).

\end{document}